# Temperature Detection from Images Using Smartphones

Kamrul H Foysal, Bipasha Kundu, and Jo Woon Chong

*Abstract*— Since late 2019, the global spread of COVID-19 has affected people's daily life. Temperature is an early and common symptom of Covid. Therefore, a convenient and remote temperature detection method is needed. In this paper, a non-contact method for detecting body temperature is proposed. Our developed algorithm based on blackbody radiation calculates the body temperature of a user-selected area from an obtained image. The findings were confirmed using a FLIR Thermal Camera with an accuracy of 97%.

*Clinical Relevance*—Our proposed method provides a remote and convenient solution in detecting temperature of specific body parts using a smartphone.

## I. INTRODUCTION

Approximately 464 million people have been infected worldwide due to covid according to World Health Organization. A common symptom of COVID-19 is fever with high body temperature, which may lead to death [1] . since the testing of COVID in a contact-based way can be inconvenient, costly, and time consuming. Especially, this contact-based way can spread COVID-19 more. Non-contact infrared thermometer sensors, devices and thermal imaging systems have been widely used to measure a person's temperature [2] However, these cannot differentiate the temperature between specific body parts of humans [2], resulting in inaccurate temperature measurements. In this paper, we propose a novel smartphone camera-based temperature detection method which estimates the temperature of any specific body part of a human using only a smartphone camera.

## II. METHODS

Pseudocolor image generation such as Jet color space has been used for thermal image color-mapping [3]. Since the temperature relationship is not linear, the estimation of temperature from the thermal images is difficult. The temperature is calculated using the obtained RGB image by creating a pseudocolor space. Then, a linear relationship between temperature and color intensity was established. According to blackbody radiation theory, a low temperature increases the visibility of red light (700nm). In contrast, a high temperature increases the visibility of blue light (490nm) and the dominant color changes with the temperature. Eq. (1) shows the calculation of spectral radiance density ($B_v(T)$) of the red light with the temperature $T$ in absolute temperature unit.

$$B_{red,v}(T_{abs}) = \frac{2v^2}{c^2} \frac{hv}{e^{hv/kT} - 1} \quad (1)$$

where $h$ is the Planck constant, $k$ is Boltzmann constant, and $v$ is the frequency of the red light. In the pseudo color space, each image pixel value represents a specific temperature data point (corresponds to $B_v(T)$). These data points are assigned a unique color or shade based on their value. As the temperature changes, the pixel value changes accordingly. As for lower temperatures (<800K), the red color channel is known to be effective in estimating temperature, and pixel values of pseudo color images are proportional to the red channel value of the original RGB image. A linear relationship (see Eq. (2)) is established between the temperature ($T_{cel}$) and the pixel intensities (*I*) by using the grayscale image which is obtained from the pixel values of red channel image. The temperature detection algorithm described above can approximate the temperature of different points in degrees Celsius.

$$T_{cel} = T_{low} + (T_{high} - T_{low}) \times \frac{(I_{max} - I_{min})}{I_{max}} \quad (2)$$

where $T_{cel}$ is the temperature of the location, $T_{low}$ and $T_{high}$ are low and high temperatures, respectively, and $I_{min}$ and $I_{max}$ are lowest and highest intensities of the image, respectively.

## III. RESULTS

Following the IRB from Texas Tech University (IRB#: 2019_150), the smartphone is used to acquire an image as shown in Fig. 2(a), then our algorithm estimates the temperature of a subject's body. Figs. 2(a) and 2(b) are the RGB and pseudocolor images acquired from a subject, respectively. Fig. 2(c) shows the zoom-in version of the Fig. 2 (b) and our algorithm estimates the temperature of the forehead part from Fig. 2(c), using Eq. (2) Here, the temperature is estimated to 33.8˚C. A FLIR Thermal camera (see Figs. 2d and 2e) is used to validate the data. The obtained accuracy is 97%.

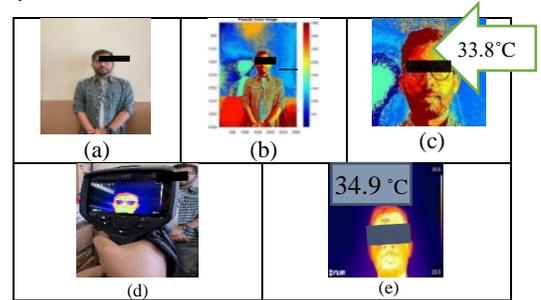

Figure 1. Temperature Detection Algorithm Results (a) RGB Image, (b) pseudo color Image, (c) ROI, (d) Taking an image with a thermal camera (34.6˚C), (e) Validation of temperature using a thermal camera (34.9˚ C)

## IV. CONCLUSION

We have developed a smartphone-based temperature estimation method which can rapidly and conveniently estimate human temperature, which is expected to reduce human encounters, and risk of spreading the virus.

*Research supported by NSF. KHF, BK & JWC are with the Texas Tech University, Lubbock, TX, USA, e-mail: bkundu@ttu.edu.